\documentclass[sigconf]{acmart}

\acmConference[]{}{}{}
\acmYear{}
\copyrightyear{}
\acmPrice{}
\acmDOI{}
\acmISBN{}
\setcopyright{none}

\usepackage[nameinlink]{cleveref}
\usepackage{stfloats}
\usepackage{xcolor}
\usepackage{xspace}

\definecolor{GREEN}{rgb}{0.0,0.6,0.0}
\definecolor{BLUE}{rgb}{0.0,0.2,0.7}
\definecolor{GOLD}{rgb}{0.6,0.6,0.0}
\definecolor{CYAN}{rgb}{0.0,0.5,0.5}
\definecolor{PURPLE}{rgb}{0.5,0.0,0.5}
\definecolor{RED}{rgb}{0.7,0.0,0.0}
\definecolor{GRAY}{gray}{0.5}
\definecolor{LIGHTGRAY}{HTML}{DDDDDD}
\definecolor{REVISIONRED}{HTML}{9e092f}
\definecolor{REVISIONGREEN}{HTML}{427a11}
\definecolor{REVISIONGOLD}{HTML}{806b06}

\definecolor{GOLD}{HTML}{FFE052}

\usepackage{enumitem}
\setlist[itemize]{leftmargin=*}
\setlist[enumerate]{leftmargin=*,label=\arabic*.}

\usepackage{listings}
\usepackage{xcolor}

\lstdefinelanguage{json}{
    basicstyle=\ttfamily\small,
    numbers=left,
    numberstyle=\tiny\color{gray},
    stepnumber=1,
    numbersep=8pt,
    showstringspaces=false,
    breaklines=true,
    frame=none,
    backgroundcolor=\color{gray!5},
    framesep=4pt,
    xleftmargin=10pt,
    xrightmargin=4pt,
    fillcolor=\color{gray!5},
    rulecolor=\color{gray!5},
    stringstyle=\color{teal},
    literate=
      *{0}{{{\color{orange}0}}}{1}
       {1}{{{\color{orange}1}}}{1}
       {2}{{{\color{orange}2}}}{1}
       {3}{{{\color{orange}3}}}{1}
       {4}{{{\color{orange}4}}}{1}
       {5}{{{\color{orange}5}}}{1}
       {6}{{{\color{orange}6}}}{1}
       {7}{{{\color{orange}7}}}{1}
       {8}{{{\color{orange}8}}}{1}
       {9}{{{\color{orange}9}}}{1}
       {:}{{{\color{black}:}}}{1}
       {,}{{{\color{black},}}}{1}
       {\{}{{{\color{black}{\{}}}}{1}
       {\}}{{{\color{black}{\}}}}}{1}
       {[}{{{\color{black}{[}}}}{1}
       {]}{{{\color{black}{]}}}}{1},
}
\begin{document}
% this seems to help remove words going beyond the margin
\tolerance=400 

% The title should be short but descriptive, like a mini abstract. 
% If possible, come up with a catchy name for your project and use it as part of the title.
% can include a short version of the title for the running header (in square brackets)
%\title[AInimation]{AInimation: Using Animated Transitions\\to Better Understand AI-Generated Responses}
\title[AInimation]{AInimation: Animating from Prompt to AI-Generated Responses}

\author{Jiaqi Wu}
\affiliation{%
  \institution{Université de Montréal}
  \city{Montreal}
  \state{Quebec}
  \country{Canada}
  }
\email{jiaqi.wu@umontreal.ca}

\author{Damien Masson}
\affiliation{%
  \institution{Université de Montréal}
  \city{Montreal}
  \state{Quebec}
  \country{Canada}
}
\email{damien.masson@umontreal.ca}

%\renewcommand{\shortauthors}{}

\begin{abstract}
% maximum 150 word abstract (aim for ~120)

We explore the use of animated transitions between a prompt and an AI-generated response. After reviewing 800 examples of prompts and responses, we devise a taxonomy of animated transitions for multimodal text- and image-generative models. The proposed animations include translating and morphing elements of the prompt to their final location in the response; highlighting modifications such as fixed typos; overlaying structural requirements to verify them; and displaying how a model understands references. A study shows that adding animated transitions helps users review the response: participants performed 43\% better at locating elements in the response; 153\% better at identifying changes; and 20\% better at verifying the prompt was correctly interpreted. Our work applies to all software that integrates AI and shows that well-crafted, slower animations are preferable to instant AI responses.

\end{abstract}




\begin{teaserfigure}
 \includegraphics[width=\textwidth]{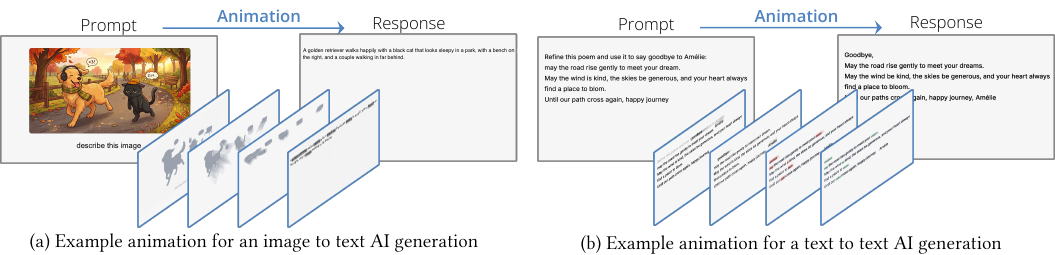}
 \caption{We design, demonstrate, and evaluate a set of animations between a prompt and a response: (a) elements of the image are morphed into the response; (b) added and removed characters are highlighted during the transition.}
 \Description{Figure for a simple example of an animated transition on the image-to-text and text-to-text task.}
 \label{fig:teaser}
\end{teaserfigure}


\maketitle
\section{Introduction}

Responses generated by AI models, such as large language models (LLMs), are typically displayed progressively, token by token. Far from being a design choice, this animation is a happy accident resulting from the underlying technology: models like the GPT family predict the next token based on previous ones~\cite{vaswaniAttentionAllYou2017, radfordImprovingLanguageUnderstanding2018}. Due to the time it takes to generate these tokens, this creates this smooth, almost word-by-word animation of the response.

Unfortunately, this animation does little to help users review the AI response, as it forces a linear read. Most changes to this animation have come from speeding up the generation thanks to improved computing speed. However, unlike computing speed, our human capacity to read, interpret and forage long written responses generated by AI will forever be bounded~\cite{perniceHowPeopleRead2014, xiaoStreamingFastSlow2025}. Previous studies have already highlighted how long LLM-generated textual answers slow users down, who either painstakingly read the whole response or skim it, missing important information~\cite{jiangGraphologueExploringLarge2023, massonDirectGPTDirectManipulation2024}.

Instead of linear animations or showing the result instantaneously, previous work showed that well-designed animations can make changes easier to follow and understand~\cite{ shanmugasundaramEffectAnimatedTransitions2008, heerAnimatedTransitionsStatistical2007, bedersonDoesAnimationHelp1999}. %Essentially, animated transitions highlight how users can carry over their knowledge to the new state. 
Such animated transitions can enhance users' comprehension of changes in data visualizations~\cite{heerAnimatedTransitionsStatistical2007, yeeAnimatedExplorationDynamic2001}, text history~\cite{chevalierUsingTextAnimated2010}, and spatial representations~\cite{bedersonDoesAnimationHelp1999, shanmugasundaramEffectAnimatedTransitions2008}. Unfortunately, these transitions have remained limited to cases where the mapping between input and output is clear, such as transitioning from a pie chart to a bar chart~\cite{heerAnimatedTransitionsStatistical2007}, or going from markup code to rendered output~\cite{dragicevicGliimpseAnimatingMarkup2011}. Using animated transitions when the input and output differ substantially, such as between a prompt and an AI response, remains an open problem.  

Inspired by these approaches, we propose to reveal the AI response by smoothly transforming parts of a prompt into corresponding parts of the response. For example, in the scenario of a user prompt, ``give me pros and cons of using GenAI''. The prompt elements ``pros'' and ``cons'' would move and morph into the corresponding parts in the response to help users track them. Similarly, when prompting ``fix the grammar of this email'', the text of the email would be moved and then modifications to it would be animated to help users identify the nature of the changes.

Our hypothesis is that animating the transition from prompt to response helps users review AI responses. We test this hypothesis by first establishing a taxonomy of animated transitions between elements of the prompt and the response. Our taxonomy is based on a review of 800 user sessions with AI models, including multimodal interactions using text and image to generate text and images. Results from three experiments show that our animated transitions resulted in a 43\% improvement in locating elements in the response, a 153\% improvement in identifying changes the AI made, and a 20\% improvement in confidently verifying that the AI correctly responded to the prompt. These results were confirmed by participants' subjective assessments, which showed that they preferred animated transitions and found them engaging.
Based on these results, we caution designers of AI systems to provide responses as quickly as possible and instead demonstrate how animated transitions can improve performance.
\section{Background and Related Work}
Most systems already animate AI responses. We review these animations and discuss other solutions to help make sense of AI responses. We conclude with the reasons animated transitions are promising.

\subsection{Current Animations of AI Responses}
Current animations of AI responses are typically due to the inner workings of the underlying AI model. As the architectures of AI models changed over the years, so did the animations. For example, early neural networks to generate images typically used Variational Autoencoders~\cite{kingmaAutoEncodingVariationalBayes2022} and Generative Adversarial Networks~\cite{goodfellowGenerativeAdversarialNets2014}. Because these models used a ``single forward pass'', they could not produce intermediate results, preventing an animation. This changed with Pixel Recurrent Neural Networks~\cite{oordPixelRecurrentNeural2016} that generate images sequentially, resulting in pixel-by-pixel animation. Today, image generation is typically powered by diffusion models~\cite{hoDenoisingDiffusionProbabilistic2020} that iteratively denoise an image. This is why tools like Midjourney produce a blurry image that becomes increasingly clear.

Text generation models have followed a similar trajectory. Early text generators like recurrent neural networks~\cite{mikolovRecurrentNeuralNetwork2010}, up until more recent transformers~\cite{vaswaniAttentionAllYou2017} generate text token by token. These tokens can be shown as they are generated, leading to this word-by-word animation in tools like ChatGPT, Gemini, or Claude. Had architectures like diffusion language models~\cite{liDiffusionLMImprovesControllable2022} been powering these tools, the animation would be a random sequence of words and characters that progressively refines into coherent sentences (e.g., see the animation of tools like Inception~\cite{InceptionNewFrontier}).

The issue is that those technology-driven animations are essentially glorified progress bars: they show that the AI is working and thus improve user experience in many ways~\cite{bhatHowDynamicVs2025}, but there is little evidence that they help users make sense of the generated response. Additionally, these animations will likely become a thing of the past: AIs like ChatGPT are becoming increasingly faster and already far exceed human reading speed~\cite{xiaoStreamingFastSlow2025}.
For these reasons, now is the time to ask the question: Should AI answers be shown as fast as possible using technology-driven animations, or would slower but well-crafted animations be preferable?

\subsection{Helping Users Make Sense of AI Responses}
Researchers have explored various methods to help users understand AI responses. For instance, one approach consists of visualizing the causal relationship between the prompt and the response. From a technical standpoint, there are many ways to recover this causal relationship, such as analyzing certain layers of the AI model~\cite{jain2019attention, wiegreffe2019attention, abnar2020quantifying, vig2019bertviz}, computing the output sensitivity~\cite{sundararajan2017axiomatic, atanasova2020diagnostic}, or slightly changing the prompt to see the impact on the output~\cite{li2016understanding, feng2018pathologies}. However, these approaches are intended for AI engineers who want to debug AI models, rather than solutions for end users.

Another approach to help users make sense of AI responses is to rethink how those responses are presented. For example, Sensecape~\cite{suhSensecapeEnablingMultilevel2023} and Graphologue~\cite{jiangGraphologueExploringLarge2023} present AI responses in a graph-based interface to structure information and facilitate exploration and understanding. Similarly, Luminate~\cite{suhLuminateStructuredGeneration2024} proposes to organize responses in a design space so users can understand what the answer is about by looking at its position along a few dimensions.

The disadvantage of these approaches is that changing the response presentation might not always be desirable nor feasible. In contrast, animated transitions are broadly applicable across AI-integrated systems because they do not occupy permanent screen real estate and adapt to a wide range of interfaces, from conversational interfaces to AI-powered drawing software.

\subsection{Animated Transitions}
Animated transitions are a sequence of visual frames (usually temporally interpolated~\cite {chevalierUsingTextAnimated2010}) to smoothly convey the relationship between one display state and another~\cite{heerAnimatedTransitionsStatistical2007, chevalierNotsoStaggeringEffectStaggered2014}. Well-crafted ones help users perceive, track, and understand the correspondences between elements across two states~\cite{dragicevicTemporalDistortionAnimated2011, liRouteFlowTrajectoryAwareAnimated2025}. Because they are temporary, they take virtually no screen real estate, making them broadly applicable~\cite{dragicevic2011gliimpse}. For these reasons, many researchers and designers have advocated their use in user interfaces~\cite{AnimationInterface1990, hudsonAnimationSupportUser1993, chevalierAnimations25Years2016}. Yet, they have remained rare due to the misconception that they ``slow down'' the interaction~\cite{hudsonAnimationSupportUser1993, bedersonDoesAnimationHelp1999} (it is often quite the contrary~\cite{shanmugasundaramEffectAnimatedTransitions2008}).

Benefits of animated transitions have been empirically verified across many domains. For example, a smooth animation of viewpoint changes helps users build a ``mental map'' of spatial information~\cite{bedersonDoesAnimationHelp1999}, reduce errors~\cite{kleinBenefitsAnimatedScrolling2005} and double user speed~\cite{shanmugasundaramEffectAnimatedTransitions2008}. Similarly, animating the transition between statistical charts (by morphing the marks from one representation to the other) was found to help track elements and estimate changes~\cite{heerAnimatedTransitionsStatistical2007}. Following these successes, researchers have investigated the parameters responsible for successful transitions, such as speed~\cite{dragicevicTemporalDistortionAnimated2011}, staggering~\cite{chevalierNotsoStaggeringEffectStaggered2014, heerAnimatedTransitionsStatistical2007}, motion~\cite{dragicevic2011gliimpse, liRouteFlowTrajectoryAwareAnimated2025}, and bundling~\cite{duTrajectoryBundlingAnimated2015, zhengFocus+contextGroupingAnimated2018, wangVectorFieldDesign2018}.

Most relevant to this paper are two animated transition techniques for textual documents.  Diffamation~\cite{chevalierUsingTextAnimated2010} animates text insertions and deletions to help identify changes when navigating between versions of a document. Gliimpse~\cite{dragicevicGliimpseAnimatingMarkup2011} enables smooth transitions between document markup code and its visual rendering by moving matched elements and interpolating them when the font differs, or text becomes an image.

While powerful, these animated transitions were designed assuming a clear relationship between states, not for complex, nondeterministic relationships that differ in semantics, modality and even presentation. 
Thus, we build on these animations and empirical results to design animated transitions for AI and hypothesize that they help users review the AI-generated content.

\section{Animating From Prompt to Response}\label{sec:goals}
Based on previous classifications of the roles of animations in user interfaces~\cite{chevalierAnimations25Years2016, AnimationInterface1990}, we identified four key design goals for animated transitions between prompt and AI responses.

\subsubsection*{G1: Locate elements of interest} When the AI is prompted to use specific elements or respond to questions, the animation should highlight where these elements and answers end up in the response. For example, when prompting ``what is the weight and size of the Eiffel Tower'', the animation should highlight where the responses for size and weight are located in the response.

\subsubsection*{G2: Estimate changes} When the AI modifies an element specified in the prompt, those changes should be highlighted such that users can tell the nature of these changes, where they occurred and how much has changed. For example, when prompting to ``fix the grammar of this email'', the animation should highlight which words were changed and how they were changed.

\subsubsection*{G3: Verify the interpretation of the prompt} The animation should surface how the prompt was interpreted, such that users can check whether this interpretation was correct and where the instructions were applied. For example, when giving requirements such as ``only five sentences'' or ``only the second sentence'', the animation should make it obvious whether these requirements were followed.

\subsubsection*{G4: Keep the user engaged} The animation should maintain users' attention, particularly when the generated answer is long and complex. Unlike the current word-by-word animation that most users tend to ignore, users should not be tempted to leave the application until the animation is complete to avoid the cognitive burden associated with context switching~\cite{rogersCostsPredictibleSwitch1995}. 

\subsection{Methodology}
We define the space of possible animations by reviewing pairs of prompt-response from real-world user sessions with AI systems. 

\subsubsection*{Dataset} To cover a large space of AI use cases, we considered four categories: textual prompts generating textual responses (text-to-text), prompts composed of text and images resulting in textual responses (image-to-text), prompts composed of text and images resulting in text and image responses (image-to-image), and textual prompts resulting in text and image responses (text-to-image). For each category, we sampled 200 pairs of prompt and response from four datasets: WildChat~\cite{zhao2024wildchat}, JourneyDB~\cite{sun2023journeydb}, Visionarena~\cite{chou2025visionarena}, HumanEdit~\cite{bai2024humanedit}. Pairs for which we could not understand the prompt (due to language, typos, or unclear phrasing) were removed and replaced. In total, we reviewed 800 pairs of prompts and responses.

\subsubsection*{Review} One author reviewed all the pairs of prompts and responses to get familiar with the data. Then, for each pair of prompt and response, the author highlighted related elements in both the prompt and response and, for each, considered different animated transitions that might be useful. The author then compared and reviewed all the animations across modalities to generalize and update candidate animations based on their similarities and differences.  

\subsubsection*{Discussion} All authors then discussed the results over hour-long meetings. Edge cases were discussed, similar animations were merged, and results were consolidated into a taxonomy.

\subsubsection*{Validation} We sampled another 100 pairs of prompt and response from each dataset and verified that each fit within our taxonomy. 

\subsection{Defining the Elements to Animate}
An animated transition requires a mapping between the elements of a source and target ``display''. Prior research focused on cases where this mapping is clear because the same elements appear in the source and target displays~\cite{heerAnimatedTransitionsStatistical2007, dragicevic2011gliimpse}. In contrast, the mapping between prompt and response is unclear because elements vary in granularity, modality (e.g., text becomes an image), and semantics.

In the context of animated transitions for prompt-response, we define an element as a semantically meaningful unit that can be a text passage (e.g., a word, an expression, a paragraph) or part of an image (e.g., a dog, the background). For a mapping between a prompt and a response element to exist, there must be a causal relationship (i.e., removing the element from the prompt would remove the corresponding element in the response).

In the simplest case, similar elements appear in both prompt and response, such as asking an LLM to write a story involving Alice: ``Alice'' is likely to appear in the response. To determine the granularity of elements, we use the ``largest common denominator'' between the prompt and the response. For example, assuming the expression ``brown angry dog'' in the prompt, if the response refers to it as ``dog'', ``brown dog'', and ``angry dog'', then it is the largest common denominator that will be animated (here, ``dog''). Elements also need to be semantically related.  Consider a prompt containing ``Bob has a brown cat'' and ``Bob moves like a cat''. If the response refers to ``Bob's cat'' it should not match with ``moves like a cat''.

The most complicated cases occur when there is a strong causal relationship, but the elements bear no resemblance (e.g., asking the LLM to generate something). In this case, we still use the largest common denominator, except the match is done semantically. For example, if generating an image of a ``brown angry dog'', that whole expression will be matched to the dog in the image, as long as it matches the description.

\subsection{Taxonomy of Animations}\label{sec:taxonomy}

\begin{figure*}[t]
    \includegraphics[width=\textwidth]{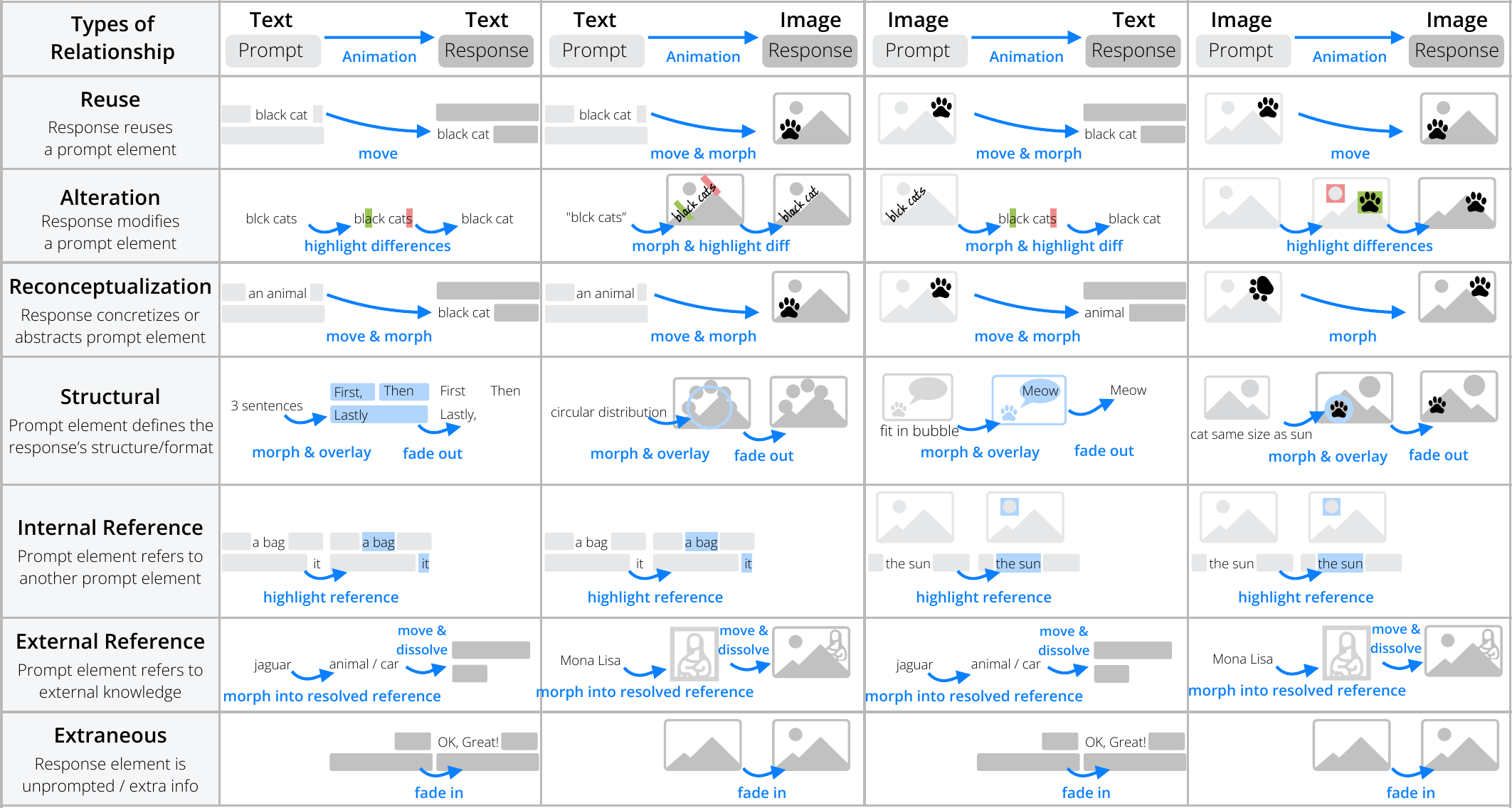}
    \caption{Our complete taxonomy of animated transitions across considered modalities and types of relationship between prompt and response. Light gray is the prompt, dark gray the response, and blue the animation.}
    \label{fig:designspace}
    \Description{TODO}
\end{figure*}

Through our review of prompts and responses, we identified seven types of relationships between elements from the prompt and response. For each, we devised what an ideal animation could look like, such that it fulfills our design goals. The full taxonomy and associated animations across modalities is shown in \Cref{fig:designspace}.

\subsubsection*{\textsc{Reuse}} Elements from the prompt are reused in the response. For example, when asking an LLM to write a story about Alice and Bob, it is likely that the response will reuse the names ``Alice'' and ``Bob''. We suggest a translation animation where ``Alice'' from the prompt moves to its final location in the response. In case ``Alice'' appears multiple times, the word Alice is duplicated and distributed across the response. Similarly, when asking to generate an image with a ``bird'', the word bird should move and morph into the bird in the image. This allows to keep track of elements and build a ``spatial memory'' of the response to more quickly skim it (G1).

\begin{figure}[H]
    \includegraphics[width=.475\textwidth]{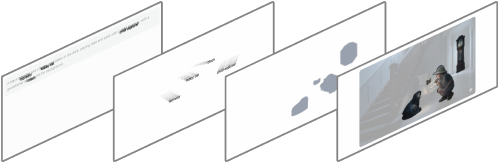}
    \caption{\textsc{reuse} animation for text to image generation. The last step shows the beginning of \textsc{extraneous} animation.}
    \label{fig:reuse}
    \Description{}
\end{figure}

\subsubsection*{\textsc{Alteration}} Elements from the prompt are reused but with slight modifications. A frequent case is when asking an LLM to ``fix the grammar of the following text''. The text will be reused but altered. We propose to animate this process through a text animation following the guidelines set by \citet{chevalierUsingTextAnimated2010}: the deleted characters fade out with a red background and the added characters fade in with a green background. We adapt this animation for images as well, such that removed objects flash red, added ones flash green, and other altered pixels flash red, then green. This allows users to check AI modifications and detect spurious ones (G2).

\begin{figure}[H]
    \includegraphics[width=.475\textwidth]{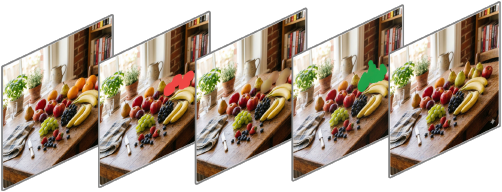}
    \caption{\textsc{alteration} for image to image generation.}
    \label{fig:t2i}
    \Description{}
\end{figure}

\subsubsection*{\textsc{Reconceptualization}} Elements from the prompt are either concretized or abstracted. For example, when an LLM is asked to give an example or respond to a question, the generated response is a concretization. Inversely, the response might abstract concepts from the prompt: when asking to write a story about cats and dogs, the response might use ``the animals''. Similar to \textsc{Reuse}, we propose to animate this relationship through a translation: Elements from the prompt will move and morph into the corresponding element in the response. This should help users build a spatial memory of the response and point them to the information of interest (G1).

\begin{figure}[H]
    \includegraphics[width=.475\textwidth]{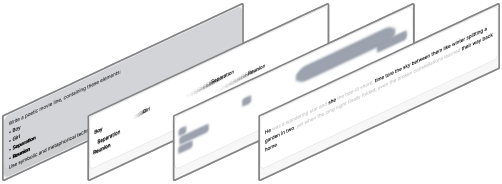}
    \caption{Example \textsc{reconceptualization} animation. The last step shows the beginning of \textsc{extraneous} animation.}
    \label{fig:rerconcep}
    \Description{}
\end{figure}

\subsubsection*{\textsc{Structural}}  Elements from the prompt decide the structure and format of the response. For example, a prompt might ask for a specific length or a specific shape (e.g., bullet points). Similarly, when generating or modifying an image, a prompt might provide relational instructions (e.g., ``not bigger than the dog'') or specify the ratio of the image (e.g., a 16:9 photo). We propose animating these expressions by morphing them into the required shape or format (e.g., a contour frame or a shape highlight), overlaying them on the final answer, and fading them out after some time. 
This gives an easier way to check the correct application of requirements (G3).

\begin{figure}[H]
    \includegraphics[width=.475\textwidth]{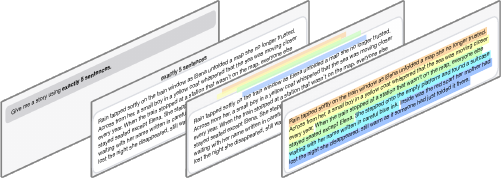}
    \caption{\textsc{structural} animation for text-to-text generation. The structural overlay will eventually fade out.}
    \label{fig:rerconcep}
    \Description{}
\end{figure}

\subsubsection*{\textsc{Internal Reference}} Elements from the prompt refer to other elements from the prompt. For example, pronouns such as ``he'' or ``these'' refer to previously mentioned names or expressions. Some of these pronouns might be an ambiguous reference. We suggest highlighting the referee and the referent in the prompt with the same colour. A similar animation is possible in multimodal tasks, when both text and image are used as the prompt, such as highlighting ``the dog'' in text and the corresponding element on the image. This prompt-only animation is technically not a transition but allows users to verify the interpretation (G3).

\subsubsection*{\textsc{External Reference}} Elements from the prompt might refer to concepts from the world. These concepts can only be understood by relying on the AI's knowledge of the world. We propose to morph these external concepts into intermediary, concretized ideas (e.g., text specifying the hidden conditions are contained in the prompt element, or a visualization of a specific art style required by the prompt element) to highlight what the AI is understanding, without affecting the generation of the response and other animations. After staying enough time for users to review those ideas, they will move and dissolve into the generated response text, indicating the influence of AI's interpretation on the response. This is meant to help users verify the correct interpretation of the prompt (G3), especially when using ambiguous concepts such as ``jaguar'', which could refer to either a car brand or the animal.

\begin{figure}[H]
    \includegraphics[width=.475\textwidth]{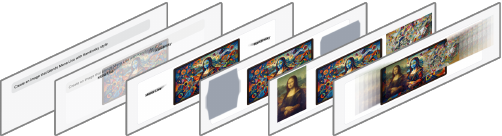}
    \caption{\textsc{external reference} animation. Prompt elements "mona lisa" and "style of Wassily Kandinsky" morph into images and then move and dissolve into the response image.} 
    \Description{}
\end{figure}

\subsubsection*{\textsc{Extraneous}} For all elements in the response that do not have a clear relationship to the prompt, we propose to show them through a fade-in animation at the very end. For example, LLMs such as ChatGPT tend to add extra information, such as encouragements (``great idea!'') or follow-up questions. By animating them differently from the rest and at the end, users can focus first on the elements directly relevant to their prompt (G1).

\subsection{Coordinating and Timing Animations}\label{sec:timeline}
Complex prompts might yield many different animations. To remain coherent and intelligible, some animations might need to start and end before others, while others might need to be distorted. Below, we highlight two common cases and discuss timing issues.

\begin{figure}[H]
    \includegraphics[width=.475\textwidth, trim=0 40px 72px 0, clip]{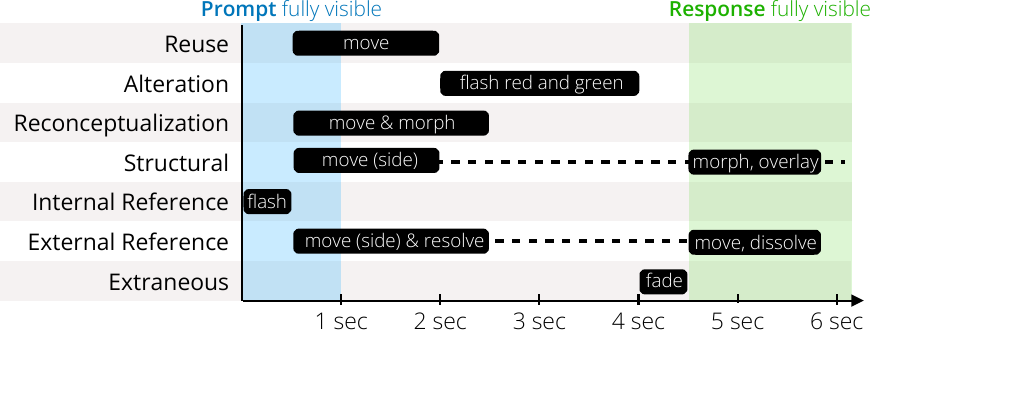}
    \caption{Timeline of the animations. Horizontal axis is the evolution of the animation, from prompt to response. Dashed line indicates when an animation is idle.}
    \label{fig:timeline}
    \Description{}
\end{figure}

\subsubsection{Coordinating Dependent Animations}
Some animations depend on others and can only be executed after them. \Cref{fig:timeline} shows when each animation should be started and ended, assuming the prompt involves all possible animations (a case that is unlikely in practice). For example, the \textsc{internal reference} animation occurs only at the beginning, since it only appears in the prompt. Other animations play in two parts. For example, the \textsc{structural} animation first extracts the structural element from the prompt and later overlays it on the final response. Thus, the overlaying animation needs to happen at the very end, once the response is fully generated. Same for the \textsc{external reference} animation, which requires extracting the reference, resolving it on the side, and showing where it applies in the response at the end.

\subsubsection{Coordinating Simultaneous Animations}
A known limitation of animated transitions is that many simultaneously animated elements might overwhelm users~\cite{simonsChangeBlindness1997}. This is particularly true for animations such as \textsc{reuse}, where the moving elements might overlap. To ensure the animation remains intelligible, we propose two strategies: like in Gliimpse~\cite{dragicevic2011gliimpse}, elements can follow curved trajectories to minimize overlap. When the number of animated elements is large, this can be combined with the bundling strategy proposed by \citet{duTrajectoryBundlingAnimated2015}: elements sharing similar paths and spatial proximity are bundled so that users focus only on one ``bundle''. Alternatively, more complex bundling strategies might be employed depending on situations~\cite{wangVectorFieldDesign2018, zhengFocus+contextGroupingAnimated2018, liRouteFlowTrajectoryAwareAnimated2025}.

\subsubsection{Timing Animations}
Different animations have different minimum durations. Prior work showed that one second was enough to track movement and that animations without it could be even slower~\cite{robertsonAnimatedVisualizationMultiple2002}. 
\Cref{fig:timeline} shows the minimum acceptable duration according to our tests and prior work~\cite{robertsonAnimatedVisualizationMultiple2002, heerAnimatedTransitionsStatistical2007, chevalier2010using}. For example,  \textsc{reuse} lasts 1.5 seconds, but \textsc{reconceptualization} lasts 2 seconds to take into account the morphing. 

In the worst-case scenario, where all animations are playing, the animated transition would last about 6 seconds. This is rarely the case in practice because prompts usually use a subset. For example, a common case is \textsc{reuse} and \textsc{extraneous}, which would only take 2 seconds to complete. Designers can also decide not to use all the animations, depending on context. 
Inversely, the duration of each animation could be extended based on the complexity of the task and the number of elements, although we caution designers as slow animations can be disliked~\cite{baudischPhosphorExplainingTransitions2006}.

All animations with translations (\textsc{reuse}, \textsc{reconceptualization}, \textsc{structural}, and \textsc{external reference}) are temporally distorted using ``slow-in'' and ``slow-out'', which has been shown to help track elements~\cite{dragicevicTemporalDistortionAnimated2011}. That is, elements gradually accelerate as they begin their motion, then slow down as they approach their destination.

\section{Technical Feasibility}
Given a prompt and a response, generating the animation requires: 1) identifying \textit{what} pairs of elements from prompt and response should be animated; 2) for each pair, deciding \textit{how} it should be animated (i.e., which animation from \cref{sec:taxonomy} to use); and 3) playing the animations based on these information.

For this last step, we implemented a system in React that plays animations based on a prompt, a response, and an animation specification. The animation can then be configured to play at a certain speed.  The system supports all the animations in our taxonomy and works for text-to-text and multimodal cases involving images, in which case the animation specification must also include image masks. Similarly, for \textsc{external reference} animation, the specification should also specify what the element should resolve to.

The first two steps, however, are more difficult to automate, as they require analyzing the prompt and the response to decide what to animate. While the primary contribution of this paper is not technical, in this section, we attempt to demonstrate the feasibility of achieving these steps automatically. Our approach is naive and should be viewed as the floor of performance one can expect. It is likely that more clever approaches, such as analyzing the model's attention layers~\cite{jain2019attention, wiegreffe2019attention}, would be more efficient and yield better accuracy. However, we believe this is beyond the scope of an HCI paper whose primary contribution is a taxonomy and results from user experiments (that we will present next).

\subsection{Pipeline and Methodology}
We implemented a simple pipeline to create animation specifications from a prompt and a response. Our pipeline relies on the GPT 5.4 model with an optimized prompt to retrieve pairs of elements to animate and their corresponding animations.

To evaluate the performance of the pipeline, we chose five representative tasks, which cover all animation types in our taxomony: 1) refining a text piece; 2) generating a text by reconceptualizing elements; 3) generating a text using specific formatting and structural constraints; 4) taking on a persona to answer a question; 5) editing a specific part of a text. For this modest technical feasibility study, we limit the evaluation to the text-to-text modality, which is the most common in practice~\cite{chatterjiHowPeopleUse} and avoids adding image segmentation to the mix. For each of the five tasks, we design 2 sets of prompts and responses. Because our pipeline involves an LLM, the results will vary every time it is run. To ensure results are not a fluke, we run the pipeline 10 times, generating 100 animation specifications. 

We evaluated the accuracy of these generated animation specifications by manually labelling them and checking whether the produced animations follow the descriptions in our taxonomy. We then measure the accuracy of a specific animation per run. Accuracy is defined as the number of accurate animations divided by the total number of animations of that type in that run. To show how reliable the pipeline is, we also calculate boostrapped 95\% confidence intervals using 10,0000 resamples of the accuracy numbers from the different runs. We use the studentized bootstrap because it is preferable at this sample size~\cite{zhuAssessingComparingAccuracy2018}.

We do not evaluate the \textsc{extraneous} animation because it does not require the pipeline, as it only animates the remaining elements. In other words, its accuracy will depend on the other animations.

\subsection{Results}

\begin{table}[h]
\caption{Accuracy of our automatic pipeline for the different animations in our taxonomy. The experiment was repeated 10 times to calculate bootstrapped confidence intervals.}
\label{tab:pipeline-accuracy}
\begin{tabular}{lcc}
\toprule
Animation & Accuracy [95\% CI] & Avg. \#\\
\midrule
\textsc{reuse}                   &  1.00 [1.00, 1.00] & 13.8 \\
\textsc{internal reference}      &  0.98 [0.96, 1.00] & 11.9 \\
\textsc{reconceptualization}     & 0.83 [0.75, 0.90] & 24.1 \\
\textsc{structural}              & 0.77 [0.71, 0.82] & 13.3 \\
\textsc{alteration}              & 0.62 [0.57, 0.65] & 12.2 \\
\textsc{external reference}      & 0.59 [0.47, 0.76] & 11.9  \\
%\hline\hline
%Total     & 0.59 [0.47, 0.76] & 11.9  \\
\bottomrule
\end{tabular}
\end{table}

\Cref{tab:pipeline-accuracy} shows the breakdown of the results. Overall, the pipeline achieves strong performance across most animations, except \textsc{alteration} and \textsc{external reference}. After an investigation, it seems that these lower results are due to the pipeline being much more willing to create animations for seemingly obvious elements than a human. The detected animations were not ``wrong'' per se, but seemed mostly pointless. For example, in one case, the pipeline decided that ``girl'' needed an \textsc{external reference}. Similarly, the pipeline reported \textsc{alteration} even when elements differed substantially, and a \textsc{reconceptualization} would appear more appropriate.

While imperfect, we believe these results demonstrate the feasibility of an automatic approach. We hope our results encourage further research in improving our simple pipeline. Note also that accuracy in this context is not all or nothing: even with lower accuracy, the accurate animations should still help. The ``inaccurate'' ones are not misleading; they are just probably unnecessary because they are animating the obvious. 
Finally, not \textit{all} the animations have to be used. Depending on context, designers can use a subset.

\section{User Experiments}
\label{Experimentation}
While there is ample evidence in support of animated transitions~\cite{chevalierAnimations25Years2016, chevalierUsingTextAnimated2010, heerAnimatedTransitionsStatistical2007}, it is unclear whether their benefits hold for prompt-response animations for which elements change drastically, sometimes even changing modalities. Thus, we collect empirical data to check whether the proposed animations fulfill the four goals we set in \cref{sec:goals}. The three successive experiment tests the first three goals. The last goal is tested through a final questionnaire.

Sixteen participants (18 to 26 years old, M = 21.9; 8 self-identified as female and 8 as male) recruited from our local community completed the 3 experiments in succession. On a 5-point scale from 1-``never'' to 5-``often'' they reported their frequency of use of AI services such as ChatGPT as Mdn=4 and having used it for text-to-text (n=16) and text-to-image (n=11) tasks before. They sat in front of a 27" Dell S2725HS monitor using a resolution of 1920 $\times$ 1080. In appreciation for their time, participants received \$15. The study was approved by the ethics board of our institution.

For all three experiments, tasks were performed with animated transitions (referred to as \textsc{animated}) and a baseline that showed the response instantaneously (\textsc{instant}). Alternative baselines were pilot-tested, including showing text word-by-word (like ChatGPT) and images through diffusion-like animations. However, the instant baseline was the strongest, with the other alternatives so weak that comparison felt moot: they did not help and instead reduced the time participants had to review the response because most tasks could only be done once the final response was shown.
The only competitive baseline was \textsc{instant}, which simulates a near-future in which computing power enables immediate answer generation. To ensure fair comparison, the exact same time was used for both conditions. That is, if an animated transition took 5 seconds to complete, then the response was shown for 5 seconds longer in the \textsc{instant} condition, which made that baseline particularly strong as participants had more time to review the answer. %As a result, participants always had more time with the final answer in the baseline. 

Prompts for all tasks were designed by taking inspiration from real prompts found during our analysis (\cref{sec:taxonomy}). All responses were pre-generated (and unaltered). The image-to-text responses are generated from ChatGPT 5.4-thinking, and text-to-text responses are generated from ChatGPT 5.2-thinking. For all image-generation tasks, Nano Banana 2 was used. All the \textsc{animated} used were generated from our prototype system. 

\subsection{Analysis}
For all experiments, the data analysis was planned, and all scripts were written prior to running the study. We use an estimation approach to analyze the data~\cite{cummingIntroductionNewStatistics2016, dragicevicFairStatisticalCommunication2016}, meaning our interpretation is based on confidence intervals of the mean difference. We still report p-values for unfamiliar readers but do not consider them. P-values are calculated using the Wilcoxon test because it is more robust to violations of normality than a t-test. Confidence intervals are calculated using a studentized bootstrap with 10,000 replications because it is the most appropriate for our study design~\cite{zhuAssessingComparingAccuracy2018, massonStatslatorInteractiveTranslation2023}.

\subsection{Experiment 1: Locating Elements of Interest}
The first experiment tasked participants with following the elements in the prompt and identifying their final location in the AI-generated response (G1). Knowing where information is located enables faster reading and understanding of text~\cite{lovelaceMemoryWordsProse1983, chunContextualCueingImplicit1998}, so we believe this task helps test the benefits of animated transitions.

Four modalities were considered to cover our taxonomy: text-to-text (generating a story with specific elements), text-to-image (generating an image using specific elements), image-to-text (explaining the joke in comics), and image-to-image (moving elements). For each, users had to track two elements. These elements could appear multiple times in the response, in which case participants had to locate all of them (maximum of 8 elements to locate). %The \textsc{reuse} transition was used for all of them.

In each trial, participants were asked to read the prompt (to simulate familiarity with it) and then click start. The two elements to track were highlighted in red and orange in the prompt. After 3 seconds, the highlights disappeared, and the transition played for 4 to 6 seconds, depending on complexity. We ensured the same duration was used for the corresponding tasks with the \textsc{instant} transition. For example, for an animated transition lasting 4 seconds, the corresponding instant transition showed the response instantaneously and for an extra 4 seconds. For the \textsc{animated} transition, \textsc{reuse} and \textsc{extraneous} were used. For the \textsc{instant} transition, the complete response showed immediately. The response was hidden 3 seconds after the transition finished, at which point participants clicked the final location of the elements. A counter showed how many elements remained to be selected, in case the elements appeared multiple times in the response. Once done, participants rated their confidence in their answer on a 5-point semantic differential scale. To prevent cheating, the mouse cursor was hidden until the selection task, and participants could not point at the screen.

The experiment followed a 2 Condition (\textsc{instant} or \textsc{animated}) $\times$ 4 Modalities within-subject design. Tasks were randomly ordered. For each modality, two tasks of roughly identical difficulty were prepared. Tasks were counterbalanced across conditions.

Dependent measures were the average confidence and the average Euclidean distance between participants' clicks and the centres of the target elements in the response. When an element appeared multiple times, participants' click locations were paired with the target elements to minimize the sum of the distances.

\subsubsection{Results}

\begin{figure}[H]
    \includegraphics[width=.475\textwidth]{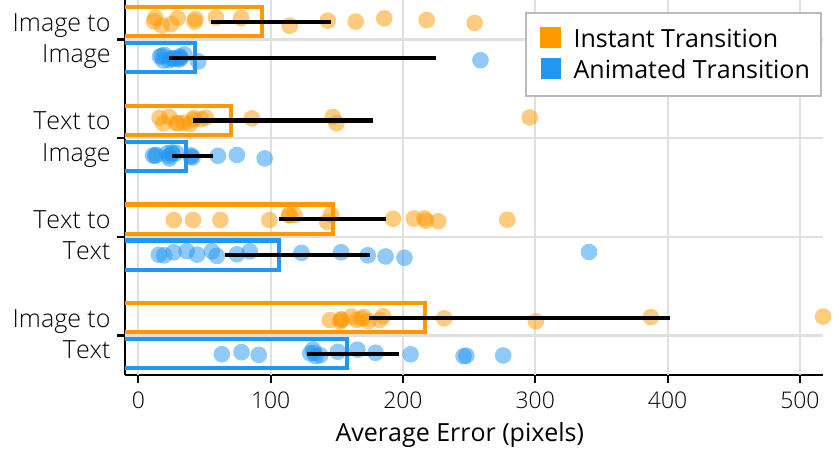}
    \caption{Average error in pixels between participants' clicks and the centre of the elements. Error bars are 95\% CIs.}
    \label{fig:distance}
    \Description{}
\end{figure}

Animated transitions reduced the average error of tracked elements by 47.5 pixels {\small (95\% CI [-73.53, -21.34], p<.0001)} corresponding to a 43\% improvement. The data distribution is shown on \cref{fig:distance}. The average error was reduced across all modalities: 
by 53.7 pixels for image-to-text {\small (M=159.21 CI [130.08 194.34] vs M=212.78 CI [173.3 372.99])};
by 48.6 for text-to-text {\small (M=108.92 CI [71.36  169.46] vs M=157.47 CI [117.17  202.08])}
by 46.7 for image-to-image {\small (M=41.89 CI [23.45 207.75] vs M=88.6 CI [53.05 140.91])}; and
by 31.3 for text-to-image {\small (M=37.44 CI [26.89  54.91] vs M=68.73 CI [41.17 180.8])}.

On a 5-point scale, participants felt more confident in their answers when they had the animated transition by .62 {\small (95\% CI [.37, .87], p<.0001)}. This was true for all modalities: 
.88 for text-to-text {\small (M=3.25 95\% CI [2.75 3.73] vs M=2.38 95\% CI [1.76 3.05])};
.69 for text-to-image {\small (M=4.56 95\% CI [4.26 4.82] vs M=3.88 95\% CI [2.99 4.42])};
.62 for image-to-image {\small (M=4.62 95\% CI [4.2 4.89] vs M=4.0 95\% CI [3.31 4.55])};
0.5 for image-to-text {\small (M=2.0 95\% CI [1.54 2.73] vs M=1.5 95\% CI [1.17 2.05])}.

\subsection{Experiment 2: Estimating Changes}
The second experiment tasked participants with estimating the changes the AI made to a text or an image provided in a prompt (G2). Modifying a text or an image is a main use case for AI~\cite{chatterjiHowPeopleUse, liValueBenefitsConcerns2024}, with users interested in checking these changes~\cite{leeImpactGenerativeAI2025}. Our hypothesis is that animated transitions help users in this process.

As in the first experiment, we considered four modalities: text-to-text (correcting the grammar), text-to-image (generating an image with text to be corrected), image-to-text (fixing the text of an image), and image-to-image (removing and adding elements). For each, participants had to estimate how many elements (images) or words (text) were removed and how many were added. 

Trials went similarly to the first experiment: they had time to read the prompt, see the animated or instant transition, and the response would be hidden 3 seconds after the transition. One difference is that participants had to pay attention to the changes to only one area of the image or the prompt. For both conditions, the transition lasted the same amount of time. After the response was hidden, they had to pick a number of elements or words that were removed and added on a scale from 0 to 10. They also had to rate their confidence in their answers on a 5-point semantic differential scale.

Dependent measures were participants' confidence and the estimation error. The estimation error was calculated as the sum of absolute errors for added and removed elements.

\subsubsection{Results}

\begin{figure}[H]
    \includegraphics[width=.475\textwidth]{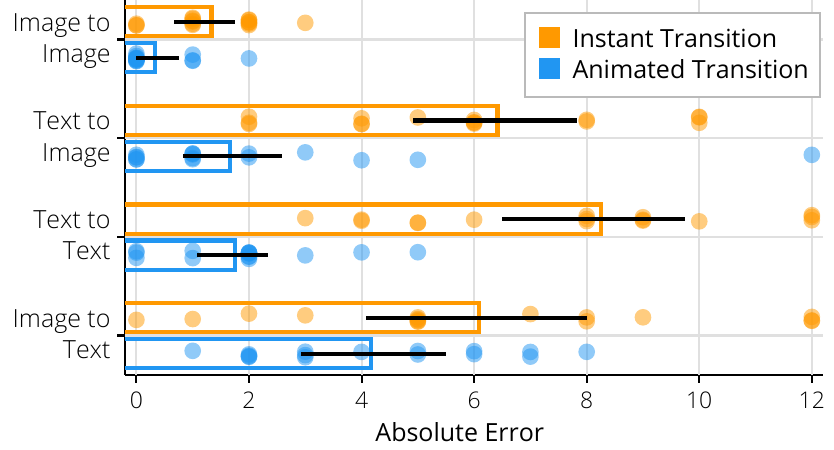}
    \caption{Absolute error when estimating changes. Each dot is a participant. Error bars represent 95\% CI.}
    \label{fig:estimation}
    \Description{}
\end{figure}

With animated transitions, participants made 3.17 fewer estimation errors {\small (95\% CI [-4.28, -2.03], p<.0001)} corresponding to a 152\% improvement. \Cref{fig:estimation} shows the distribution of the data across all four modalities. The average absolute error was systematically lower across all modalities, but the effect was largest for text-to-text:
by 5.88 for text-to-text {\small (M=1.88 95\% CI [1.26, 2.66] vs M=7.75 95\% CI [6.17, 9.23])};
by 3.75 for text-to-image {\small (M=2.06 95\% CI [.94, 5.46] vs M=5.81 95\% CI [4.48, 7.26])};
by 2.06 for image-to-text {\small (M=4.12 95\% CI [3.08, 5.35] vs M=6.19 95\% CI [4.31, 8.24])}; and
by 1.0 for image-to-image {\small (M=.31 95\% CI [.08, .89] vs M=1.31 95\% CI [.86, 1.77])}.

Participants rated themselves 1.08 points more confident in their estimations when having the animated transitions {\small (95\% CI [.79, 1.37], p<.0001)}. In particular, they were more confident across all modalities:
by 1.44 for image-to-text {\small (M=3.38 95\% CI [2.68, 3.92] vs M=1.94 95\% CI [1.43, 2.77])};
by 1.22 for text-to-image {\small (M=3.31, 95\% CI [2.71, 3.8] vs M=2.09, 95\% CI [1.65, 2.62])};
by 1.0 for image-to-image {\small (M=4.34, 95\% CI [3.98, 4.63] vs M=3.34, 95\% CI [2.71, 3.87])}; and
by .66 for text-to-image {\small (M=3.22, 95\% CI [2.65, 3.71] vs M=2.56, 95\% CI [1.93, 3.14])}.

\subsection{Experiment 3: Verifying Interpretation}
The third experiment tasked participants with rating their confidence that an instruction in a prompt was correctly interpreted and followed (G3).  Because AI is nondeterministic and known to ignore orders~\cite{zamfirescu-pereiraWhyJohnnyCant2023}, users must constantly check whether their prompts are correctly interpreted. We believe that animated transitions can help this process, which is what this task tests.

To cover the remaining transition animations in our taxonomy (\cref{sec:taxonomy}), we limited tasks to the text-to-text modality (which is the most common in practice~\cite{chatterjiHowPeopleUse}) so we could vary tasks without increasing the study duration. For each prompt, participants had to check the correct application of a specific instruction, such as only modifying a specific sentence (\textsc{reuse} and \textsc{alteration}), writing a story that includes specific elements (\textsc{reconceptualization} and \textsc{extraneous}),  taking on a persona to answer a question (\textsc{external reference}, \textsc{reuse}, and \textsc{extraneous}), and generating a story using a specific number of sentences (\textsc{structural}).

Trials went similarly to the previous two experiments, except that participants had to check the highlighted instruction in the prompt. The time for \textsc{animated} transitions followed our recommendation in \cref{sec:timeline} (4 to 6 seconds, depending on the animation). Like the previous two experiments, for both
\textsc{instant} and \textsc{animated} conditions, the transition lasted the same amount of time. The response is then hidden 3 seconds after the transition finishes, at which point participants rated how confident they are that the highlighted instruction was correctly applied.

While all responses perfectly followed the instructions in the prompt, we told participants that responses were generated by ``an imperfect AI'' that might not follow instructions. This was done to combat over-trust and force them to always check generated responses. Participants' responses thus reflected their confidence based on their assessment, rather than their general trust in AI.

\subsubsection{Results}
Overall, there is strong evidence that the animated transitions helped verify that the AI correctly interpreted the prompt. On a 5-point scale, participants were .62 more confident that the AI was correct with animated transitions {\small (95\% CI [.18, 1.07], p=.014)} corresponding to a 20\% improvement. However, the evidence and effect sizes varied across animations. There is good evidence for \textsc{alteration} {\small (by 1.5, 95\% CI [.46, 2.38], p=.014)}, \textsc{reconceptualization} {\small (by .88 (95\% CI [-.15, 1.51], p=.048)}, and \textsc{structural} {\small (by .75, 95\% CI [.17, 1.99], p=.047)}. Results were inconclusive for \textsc{external reference} {\small (95\% CI [-1.52, .64], p=.265)}.

\subsection{Subjective Feedback}
Participants responded to statements on a 5-point Likert scale at the most opportune times, ensuring each statement pertained to the tasks they had just experienced. After each condition in the first experiment, they rated how easy it was to track elements (G1). After each condition in the second experiment, they rated how easy it was to estimate changes (G2). After each condition in the third experiment, they rated how easy it was to check that the prompt was correctly interpreted (G3). At the end, they rated whether the animations were engaging (G4) and easy to understand. Participants then engaged in a brief semi-structured interview about their strategies and experience with the animated transitions.

\subsubsection{Likert Results}
\begin{figure*}[t]
    \includegraphics[width=\textwidth]{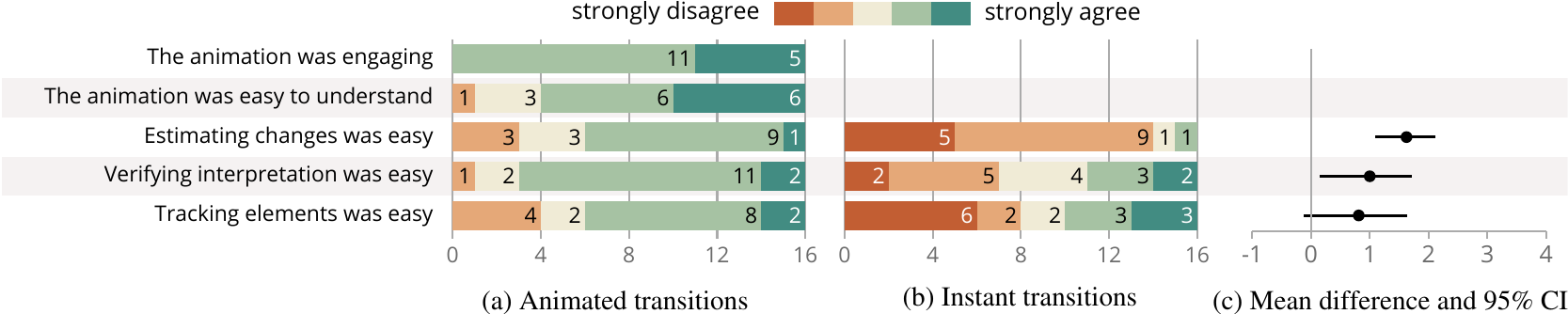}
    \caption{Responses to likert statements during the three experiments for (a) animated transitions and (b) baseline (instant transitions). Error bars represent the studentized bootstrapped 95\% CI for mean differences.}
    \label{fig:likert}
    \Description{TODO}
\end{figure*}

All 5-point scale statements were rated in favour of animated transitions. \Cref{fig:likert} shows the breakdown of the results. There is strong evidence that animated transitions were better when participants rated  ``Estimating changes was easy'' {\small (by 1.62 95\% CI [1.1, 2.12], p=.001)} and ``Verifying interpretation was easy'' {\small (by 1.0 95\% CI [.15, 1.72], p=.024)}. The evidence was weaker for ``Tracking elements was easy'' {\small (by .81 95\% CI [-.12, 1.64], p=.064)}.

When asked specifically about the animations across the whole study, participants were mostly positive. They agreed that ``The animation was engaging'' (Mdn=4 95\% [4.09 4.57]) and that ``The animation was easy to understand'' (Mdn=4 95\% [3.52 4.49]).

\subsubsection{Semi-structured Interview Results}

Overall, participants' comments confirmed the quantitative results. They all expressed a preference for the animated transitions and mentioned how they made them \textit{"more confident" } (P2) and made the tasks \textit{``much easier''} (P15) and \textit{``faster''} (P8). Part of the reason is that they could ``focus on the key changes in the response'' (P7), that \textit{``it was easy to understand if it's following the instructions well''} (P11), and that the animation provided \textit{``extra information''} (P3).

Participants also mentioned how these animations could help them in their daily tasks. For instance, P8 mentioned that the animation would be useful when they \textit{``debug code and [they] want to know what change it made to that code''} and P15 said that \textit{``[they] use AI to correct [their] text [...] and sometimes [they] don't know what it changed''}. P7 mentioned that they would be interested to use the animations to \textit{"identify the source of the response"} and compare different models to see \textit{``is it always moving object like that?''}.

Beyond our research questions, participants also appreciated that the animations were \textit{``entertaining''} (P10) and made \textit{``your text come to life''}. It also gave them the impression of seeing how the AI operates internally. For instance, P2 mentioned the animated transition \textit{"more fun to wait [...] It's interesting to see how it can create a whole block of text [...] It's like interesting to see like a bit of behind the scenes"} (P2).

While generally positive, participants did mention some issues. For instance, P10 mentioned that \textit{"[animations] can be a little distracting''}. Several participants also noted that the most useful animations involved fewer elements. And P6 and P13 both wanted slower animations. Some participants also expressed that the animated transitions were most useful for some specific tasks. For instance, P3 mentioned animations are \textit{``definitely more helpful in like the image part of the tasks''}. In contrast, P11 thought that even without the animated transitions, \textit{``when it was with an image it was really easy [to track elements in experiment 1]''}.

\section{Discussion}
Before returning to our research questions, we want to highlight the most surprising result of our experiments: animations helped \textit{despite} slowing down generation and delaying when users could see the final answer.
Even when factoring in the time the animation takes, participants were not better off using that time reviewing the final answer for longer.
Based on this result, we caution designers about the current trajectory of AI tools, which appear to view instantaneous generation as the ultimate goal.
Faster AI generation is already happening. We argue that these improved speeds should be used to develop animated transitions, not to show the response as fast as possible.
In fact, research has shown that the generation speed of popular AIs like ChatGPT already exceeds human reading speed, and there is no advantage in making it faster~\cite{xiaoStreamingFastSlow2025}.

\subsection{Summary of Results}
Based on our results, we conclude that the animated transitions we propose meet the goals we defined in \cref{sec:goals} as follows.

\subsubsection*{(G1) Our animated transitions helped locate elements in the answer} Participants were 43\% better and 25\% more confident at locating elements in the AI-generated response. Subjectively, they (n=10) agreed that tracking elements was easy, while half disagreed that it was easy with instant transitions.

\subsubsection*{(G2) Our animated transitions helped estimate changes made by AI} Participants were 153\% better and 44\% more confident in estimating how many elements were removed and added in the AI answer. Subjectively, thy (n=10) agreed that estimating changes was easy, whereas they (n=14) disagreed it was easy with instant transitions.

\subsubsection*{(G3) Our animated transition helped verify that the AI correctly followed the prompt} 
Participants were 20\% better at verifying that the AI correctly followed the instructions in the prompt. Subjectively, they (n=13) agreed that verifying interpretation was easy, while they (n=7) mostly disagreed it was easy with instant transitions.

\subsubsection*{(G4) Our animated transitions were engaging and easy to understand} All participants found the animation engaging. A large majority (n=12) thought the animations were easy to understand.

\subsection{Applications}
\textit{Prompt-driven interaction}, where users enter a prompt to control the AI, is the most straightforward case for our animated transitions. For example, \textsc{reuse} could help show where elements are placed in the image when using Adobe Photoshop's ``generative fill'' feature. \textsc{alteration} could show what was modified when using the AI features of GitHub Copilot, Grammarly, and Microsoft Excel, Word, and PowerPoint. And \textsc{structural} could help verify that generated code follows the specified formatting when using Cursor.

\textit{Chat-based interfaces} such as ChatGPT or GitHub Copilot would require only slight changes to our animations. Because the prompt remains in the history, we suggest changing the translation animations so they go from the prompt to the response shown below. Another difference is that a response might be partly caused by prior interactions. Thus, animations like \textsc{reuse} might need to start from any of the prior messages and move towards the response. This indicates that the response was influenced by prior interactions.

\textit{AI Interactions that hide the prompt} could also be augmented with animations. In most cases, \textsc{alteration} could be used, even if the AI is used through a tool such as neural filters in Photoshop or the Magic Eraser in Google Photos. Textoshop is an example of how this could be done~\cite{massonTextoshopInteractionsInspired2025}. For other animations, they would need to start from the parameters and configuration options of the interface, such as the sliders that control the neural filters in Photoshop or the tone settings in Grammarly.

\subsection{Limitations}

\subsubsection{Limitations of our results} Like all studies, our results are limited by our choices of data, participants, and study design. 
Concretely, this means that our taxonomy of animated transitions might be missing important use cases in case they did not appear in the 800 examples of prompts and responses we analyzed.
Similarly, the 16 participants in our study might not be representative of the general population, potentially missing some important perspectives. For one, all our participants were young, despite age being a known factor that impacts behaviours and software use~\cite{mahmudLearningExplorationHow2020}.

\subsubsection{Study tasks might not be good proxies for real-world performance} Our study design was inspired by studies on animated transitions~\cite{heerAnimatedTransitionsStatistical2007, chevalierUsingTextAnimated2010}. Tasks were grounded in evidence that knowing where elements are speeds up reading and comprehension~\cite{chunContextualCueingImplicit1998, lovelaceMemoryWordsProse1983} and that AI users want to understand the nature of changes~\cite{liValueBenefitsConcerns2024, leeImpactGenerativeAI2025} and check that the AI followed their instructions~\cite{zamfirescu-pereiraWhyJohnnyCant2023}.
It remains that these tasks might not cover the full range of what interests users, nor do they necessarily reflect the real world perfectly. For example, the increased cognitive load from the animation might lead to fatigue and decreased performance over time. These questions will need to be answered through longitudinal studies.

\subsubsection{Our technical implementation is limited}  While the primary contribution of our paper is not technical, we demonstrated the feasibility of an automatic pipeline and a prototype system. However, our pipeline cannot be readily integrated into an AI system without adding substantial delays. There are probably many ways to make the pipeline faster, for example, by analyzing attention layers of the AI model~\cite{jain2019attention, wiegreffe2019attention, abnar2020quantifying, vig2019bertviz}. Another option is to just wait, as AI generation has been getting faster and is expected to continue this trend, making our pipeline practical. %Another challenge that these automatic implementations face is that creating high-quality animated transitions may require complex decision-making (e.g., deciding what to animate when there are too many candidate animations) that is not easy without human intervention. 
Overall, our goal is to show a possible future and motivate further research in that direction.

\subsection{Future Work}

\subsubsection{Beyond text and images} AI models are increasingly multi-modal, supporting prompts with text, images, but also audio, video, and documents. We believe our taxonomy of animated transitions could be expanded to these alternative media types. For example, when feeding an AI a PDF, the animated transition could highlight through a \textsc{reuse} or \textsc{reconceptualization} animation where it got the information from in the document. Similarly, an \textsc{internal reference} animation could help users verify that the AI understood which part of a video or audio was intended to be referenced.

\subsubsection{Adaptable animated transitions}
One result from our experiments is that animated transitions did not help equally for all tasks and modalities. This suggests that our animated transition might need to be adapted based on the situation: perhaps animations are less critical when the task is to generate an image, because our perceptual system appears powerful enough to process images faster than an animation could guide us. Or perhaps it is not worth resolving an external reference when that same reference was already made in a previous interaction. Designers might want to decide whether all animated transitions are necessary, depending on context and user preferences.

\subsubsection{Control and regular back-and-forth between prompt and response} Some participants mentioned how the animations were sometimes too fast or overwhelming. While not tested in the study, our software offers control to pause, slow down, rewind, and speed up the animation, which might help. Alternatively, the animation could be replayed, like in Gliimpse~\cite{dragicevic2011gliimpse}, where users switch between rendered and markup code with a button press. This same paradigm could be enabled by our animated transitions: users would start with a prompt, check the result via a smooth transition, then return to the prompt to further refine the result. Animated transitions reveal causal relationships, potentially helping users know how to edit their prompts. This would shift interaction from successive prompting to refining the initial prompt.

\section{Conclusion}
We devised, technically evaluated, and empirically validated a taxonomy of animated transitions for AI. The animated transitions consist of translations, animations of changes, and morphing of elements between the prompt and the response. Results from three experiments showed that our animations helped participants locate elements in the answer, estimate the changes AI made, and verify the correct interpretation of the prompt. Subjectively, the animated transitions were also thought to be engaging and easy to understand.
Contrary to current trends that aim to generate responses as quickly as possible, our results show that well-crafted but slower animations might be preferable. We hope our work will inspire and guide designers of AI-integrated systems.

%% Acknowledgements 
\begin{acks}
We acknowledge the support of the Natural Sciences and Engineering Research Council of Canada (NSERC), RGPIN-2026-06392 and of IVADO and the Canada First Research Excellence Fund.
\end{acks}

\bibliographystyle{ACM-Reference-Format}
\bibliography{_references.bib, zotero_do_not_modify}

\end{document}